\documentclass[pra,aps,twocolumn,superscriptaddress,showpacs]{revtex4}
\usepackage{graphicx}
\usepackage{amsmath}

\newcommand{\eq}{\begin{equation}}
\newcommand{\fine}{\end{equation}}

\begin{document}

\title{Chain of Hardy-type local reality constraints for $n$ qubits}

\author{Sibasish Ghosh}
\email{sibasish@imsc.res.in} \affiliation{The Institute of Mathematical Sciences, C. I. T. Campus, Taramani, Chennai 600113, India}

\author{Shasanka Mohan Roy}
\email{shasanka1@yahoo.co.in} \affiliation{Homi Bhabha Centre for Science Education, TIFR, V. N. Purav Marg, Mankhurd, Mumbai - 400 088, India.}

\begin{abstract}
Non-locality without inequality  is an elegant argument introduced by L. Hardy for two qubit systems, and later
 generalised to $n$ qubits, to establish contradiction of quantum theory with local realism. Interestingly, for $n=2$
this argument is actually a corollary of Bell-type inequalities, viz. the CH-Hardy inequality involving Bell 
correlations, but for $n$ 
greater than 2 it involves $n$-particle probabilities more general than Bell-correlations. In this paper, we first
 derive a chain of completely new local realistic inequalities involving joint probabilities for $n$ qubits, and 
then, associated to each such inequality, we provide a new Hardy-type local reality constraint without inequalities.
 Quantum mechanical maximal violations of the chain of inequalities and of the associated constraints are also 
studied by deriving appropriate Cirel'son type theorems.
 These results involving joint probabilities more general than Bell correlations are expected to provide a new
 systematic tool to investigate entanglement.   
\end{abstract}


\maketitle

\section*{\large{1. Introduction}} Quantum theory has contradictions with local-realistic description of Nature. 
This was first revealed by the discovery that some quantum states violate the Bell-CHSH correlation inequalities 
for two qubits following from local realism  \cite{bell64}. The violations in fact become exponential in $n$ for 
some $n$-qubit states \cite{mermin90,roy-singh91,ardehalli92}. Quantum mechanical states also show non-locality 
without statistical inequalities as shown first by GHZ \cite{GHZ1989} for $n=3$ and later by a completely 
different argument by Hardy for two qubits \cite{hardy92}. The Hardy non-locality argument involves several 
experimental joint probabilities all of which except one are allowed to be zero quantum mechanically but not 
by local realism and have been generalised to multi-qubit systems 
\cite{clifton}. Further, there are generalisations changing the set of non-vanishing quantum probabilities 
to have more than one element \cite{cabello}. Recently, a set of all-versus-nothing proofs (which are multiparty 
generalization of the GHZ argument of ref. \cite{GHZ1989}) of nonlocality without inequalities for 
$n$ qubits, distributed among $m$ parties, is provided by Cabello and Moreno \cite{cabello2010};
they show that the so-called graph states \cite{nest2004} satisfy these nonlocality proofs. These all-versus-nothing 
proofs of nonlocality are associated to local realistic inequalities involving only correlation functions. 

Hardy-type locality arguments, although elegant, are weaker than Bell-type inequalities; e.g. 
no maximally entangled state of two qubits satisfies Hardy-type non-locality condition while it violates the 
Bell-CH inequality maximally. More specifically, corresponding to each Hardy-type locality constraint, 
there is a stronger constraint: a generalised  CH (which we call from now on as `CH-Hardy') \cite{clauser} local realistic 
inequality homogenous in the joint probabilities (see, for example, \cite{mermin, roy} for $n=2$ and 
\cite{cereceda} for arbitrary $n$). 

Here we report our discovery that for $n$ two-level systems, the usual CH-Hardy inequality for any $n \geq 3$ 
is just one out of a whole chain of CH-Hardy type local realistic inequalities. 
A qualitative importance of the search for new Hardy-type arguments  and 
corresponding CH-Hardy inequalities on $n$-qubit probabilities ($n \geq 3$) is that they constitute 
tests for entanglement which cannot be derived from $n$-particle Bell correlation inequalities. Although each 
$n$-th order correlation can be written in terms of $n$-qubit probabilities, each $n$-qubit probability cannot 
be written in terms of $n$-th order correlations alone. E.g. it is known that some generalized GHZ states 
for odd $n$ satisfy all Bell-correlation inequalities \cite{scarani-gisin} but violate the CH-Hardy inequalities 
for all $n$ \cite{cereceda}. Thus, for example, all three-qubit pure entangled states violate the CH-Hardy local 
realistic inequality ({\it i.e.}, involving joint probabilities only) for tripartite two-level systems 
\cite{sujit2010}. There exist computer algorithms to search for Bell inequalities 
involving all joint probabilities (not just Bell correlations) but they yield 53856 inequalities already 
for $n=3$ \cite{pitowsky} and are therefore difficult to use for general $n$. In contrast, here we provide a 
simple systematic method to list chains of CH-Hardy local realistic inequalities involving $n$-qubit 
joint probabilities and corresponding generalized Hardy-type non-locality arguments without inequalities. 
We then derive analogues of the Cirel'son theorem \cite{cirel'son} (for Bell-CHSH inequalities) for the present chain 
of CH-Hardy inequalities. The method is to calculate eigenvalues of the relevant 
quantum mechanical operators to deduce maximal violations of the inequalities for general $n$; we also find 
the corresponding maximal locality violations in the generalised Hardy-type argument without inequalities. 
The inequalities given here based on local reality are necessary conditions for separability; however, as in the case 
of Bell correlation inequalities \cite{roy2005} they can also be expected to 
trigger the discovery of quantum separability inequalities (on 
the combinations of probabilities occurring in the CH-Hardy chains) which are even stronger.

{\bf Notations:}  Consider an $n$-qubit EPR-Bell experiment by $n$ space-like separated observers where the $i$-th 
observer measures at random either $e_i$ or $e_i^{\prime}$ which can each take values 0 or 1. This setup is decribed 
in figure 1. The measured coincidence probabilities $P()$ are thus given by
\begin{equation}
\label{mcp}
\left\langle \prod_{i = 1}^{n} g_i \right\rangle = P\left(g_1 = 1, g_2 = 1, \ldots, g_n = 1\right),
\end{equation}
with $g_i = e_i$ or $\overline{e_i} \equiv 1 - e_i$, or $e_i^{\prime}$ or $\overline{e_i^{\prime}} \equiv 1 - e_i^{\prime}$. There are $4^n$ such coincidence probabilities, as for an example
$$\left\langle \prod_{k = 1}^{i} \left(1 - e_k\right) \prod_{k = i + 1}^{n} e_k^{\prime} \right\rangle =$$ 
\begin{equation}
\label{mcp1}
P\left(e_1 = \ldots = e_i = 0, e_{i + 1}^{\prime} = \ldots = e_n^{\prime} = 1\right).
\end{equation} 

As noted by Wigner \cite{wigner}, Local Hidden Variables (LHV) imply the existence of a joint probability distribution 
$$P(\{e\}, \{e^{\prime}\}) \equiv P\left(e_1, \ldots, e_n, e_1^{\prime}, \ldots, e_n^{\prime}\right)$$ 
in terms of which,
$$\left\langle \prod_{i = 1}^{n} g_i \right\rangle_{LHV} =$$ 
\begin{equation}
\label{lhvjp}
\sum_{e_1, e_2, \ldots, e_n, e_1^{\prime}, e_2^{\prime}, \ldots, e_n^{\prime}} \left(\prod_{i = 1}^{n} g_i\right)P\left(\{e\}, \{e^{\prime}\}\right),
\end{equation}
where the summation goes over 0 and 1 for each of the $e_i$ and $e_i^{\prime}$.

In contrast, in a quantum state with density operator $\rho$, 
\begin{equation}
\label{javqm}
\left\langle \prod_{i = 1}^{n} g_i \right\rangle_{QM} =~ {\rm Tr} \left[{\rho}\prod_{i = 1}^{n} G_i\right] \equiv \left\langle \prod_{i = 1}^{n} G_i \right\rangle_{\rho},
\end{equation}
where $G_i$'s (taken from the set of observables $E_1$, $E_2$, $\ldots$, $E_n$, $\overline{E_1}$, $\overline{E_2}$, $\ldots$, $\overline{E_n}$, $E_1^{\prime}$, $E_2^{\prime}$, $\ldots$, $E_n^{\prime}$, $\overline{E_1^{\prime}}$, $\overline{E_2^{\prime}}$, $\ldots$, $\overline{E_n^{\prime}}$, whose descriptions are given below) are the self-adjoint projection operators corresponding to the variables $g_i$, {\it i.e.}, $e_i \leftrightarrow E_i$, $e_i^{\prime} \leftrightarrow E_i^{\prime}$ with
\begin{equation}
\label{proj}
E_i = \frac{I + \vec{\sigma}^{(i)}.\vec{a}^{(i)}}{2},~ E_i^{\prime} = \frac{I + \vec{\sigma}^{(i)}.\vec{a}^{(i) {\prime}}}{2}.
\end{equation}  

\begin{widetext}

\vspace{1.2cm}
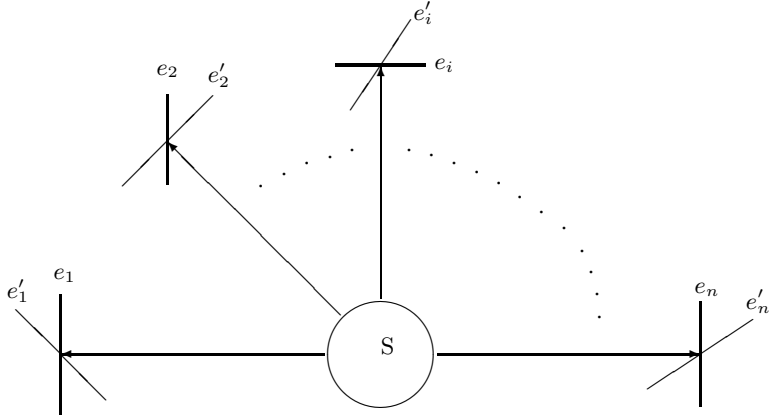
\begin{figure}[h]
\setlength{\unitlength}{0.1cm}
\centering
\begin{picture}(60, 40)
\put(15, 7){\circle{15}{S}}
\put(7.5, 7){\vector(-1, 0){35}}
\thicklines
\put(-27.5, 7){\line(0, 1){8}}
\put(-27.5, 7){\line(0, -1){8}}
\put(-29.5, 15){\makebox(5, 5){$e_1$}}
\thinlines
\put(-27.5, 7){\line(-1, 1){6}}
\put(-27.5, 7){\line(1,-1){6}}
\put(-35.7, 13){\makebox(5, 5){$e_1^{\prime}$}}
\put(9.7, 12.3){\vector(-1, 1){23}}
\thicklines
\put(-13.3, 35.6){\line(0, 1){6}}
\put(-13.3, 35.6){\line(0, -1){6}}
\put(-15.3, 42.2){\makebox(4, 5){$e_2$}}
\thinlines
\put(-13.3, 35.4){\line(1, 1){6}}
\put(-13.3, 35.4){\line(-1, -1){6}}
\put(-9, 41.7){\makebox(5, 5){$e_2^{\prime}$}}
\put(-0.9, 29.4){\circle*{0.1}}
\put(2.2, 31){\circle*{0.1}} 
\put(5.2, 32.5){\circle*{0.1}}
\put(8.3, 33.35){\circle*{0.1}}
\put(11.2, 34.2){\circle*{0.1}}
\put(15, 14.5){\vector(0, 1){31}}
\thicklines
\put(15, 45.5){\line(1, 0){6}}
\put(15, 45.5){\line(-1, 0){6}}
\put(21, 43){\makebox(5, 5){$e_i$}}
\thinlines
\put(15, 45.5){\line(2, 3){4}}
\put(15, 45.5){\line(-2, -3){4}}
\put(18.2, 50){\makebox(5, 5){$e_i^{\prime}$}}
\put(18.8, 34.2){\circle*{0.1}}
\put(21.8, 33.5){\circle*{0.1}} 
\put(24.8, 32.5){\circle*{0.1}}
\put(27.8, 31.2){\circle*{0.1}}
\put(30.7, 29.8){\circle*{0.1}}
\put(33.5, 27.9){\circle*{0.1}}
\put(36.4, 26){\circle*{0.1}}
\put(39.4, 23.8){\circle*{0.1}}
\put(41.8, 20.8){\circle*{0.1}}
\put(43, 18){\circle*{0.1}}
\put(43.8, 15){\circle*{0.1}}
\put(44.1, 12){\circle*{0.1}}
\put(22.5, 7){\vector(1, 0){35}}
\thicklines
\put(57.5, 7){\line(0, 1){7}}
\put(57.5, 7){\line(0, -1){7}}
\put(55.9, 13.2){\makebox(5, 5){$e_n$}}
\thinlines
\put(57.5, 7){\line(3, 2){7}}
\put(57.5, 7){\line(-3, -2){7}}
\put(62.7, 11.6){\makebox(5, 5){$e_n^{\prime}$}}
\end{picture}
\caption{The $n$-qubit EPR-Bell Experiment by $n$ space-like separated observers on an $n$-qubit system, evolved from the sourse $S$; the $i$-th observer measures $e_i$ or $e_i^{\prime}$, each of which can take values 0 or 1.}
\label{fig:multisetting}
\end{figure}

\end{widetext}

Here $\vec{\sigma}^{(i)}$'s are Pauli-matrices for the $i$-th qubit, $\vec{a}^{(i)}$ and $\vec{a}^{(i) {\prime}}$ are two different directions (unit vectors) corresponding to measurements of $e_i$ and $e_i^{\prime}$ respectively. Similarly, $\overline{e_i} = 1 - e_i \leftrightarrow \overline{E_i}$, $\overline{e_i^{\prime}} = 1 - e_i^{\prime} \leftrightarrow \overline{E_i^{\prime}}$, with
\begin{equation}
\label{proj1}
\overline{E_i} = \frac{I - \vec{\sigma}^{(i)}.\vec{a}^{(i)}}{2},~ \overline{E_i^{\prime}} = \frac{I - \vec{\sigma}^{(i)}.\vec{a}^{(i) {\prime}}}{2}.
\end{equation}
Thus, for example,
$$P_{QM}\left(e_1 = \ldots = e_i = 0, e_{i + 1}^{\prime} = \ldots = e_n^{\prime} = 1\right) =$$ 
\begin{equation}
\label{qmp}
\left\langle \prod_{j = 1}^{i} \overline{E_j} \prod_{k = i + 1}^{n} E_k^{\prime} \right\rangle_{\rho}.
\end{equation} 

In section 2, we discuss generalized versions of Hardy's local reality constraints on probabilities and the 
corresponding CH-Hardy inequalities. In section 3, we derive a chain of CH-Hardy type local reality inequalities. 
In section 4, we derive new Hardy-type local reality constraints without inequalities. Section 5 deals with 
quantum mechanical violation of the new Hardy-type local reality constraints without inequalities. Eigenvalues of some of 
the `Bell' operators appearing in the above-mentioned chain of local reality inequalities are derived in section 6 
and used to deduce Cirel'son-type theorems on quantum mechanical violation of these inequalities. In section 7, we 
summarise our conclusions and possible future directions.

\section*{\large{2. Hardy's local reality constraints on probabilities and corresponding CH-Hardy inequalities}} 
Suppose that the probability 
\begin{equation}
\label{ek'}
P(e_k^{\prime} = 1~ \forall k) > 0,
\end{equation}
and that
\begin{equation}
\label{eiek'}
\sum_{i = 1}^{n} P\left(e_i = 1; e_k^{\prime} = 1~ \forall k \ne i\right) = 0;
\end{equation}
then local reality implies that the probability
\begin{equation}
\label{ek}
P\left(e_k = 0~ \forall k\right) > 0,
\end{equation}
which is Hardy's local reality condition \cite{hardy92}, extended in \cite{clifton} for general $n$. The proof is 
elementary. The condition $P(e_k^{\prime} = 1~ \forall k) > 0$ implies the existence of events with all 
$e_k^{\prime} = 1$. Assuming local reality, condition (\ref{eiek'}) then requires that all the $e_i = 0$ for these events, 
and hence the condition (\ref{ek}).

Moreover, it is elementary to prove the associated CH-Hardy inequality \cite{mermin}, extended to 
general $n$ \cite{cereceda},
$$X \equiv P\left(e_k = 0~ \forall k\right) +$$
\begin{equation}
\label{hardy-CH-Hardy}
\sum_{i = 1}^{n} P\left(e_i = 1; e_k^{\prime} = 1~ \forall k \ne i\right) - P\left(e_k^{\prime} = 1~ \forall k\right)
 \ge 0,
\end{equation}
which contains the extended Hardy locality constraints, mentioned above. 

{\noindent {\bf Proof:}} The left-hand side of (\ref{hardy-CH-Hardy}) in an LHV theory is
$$X_{LHV} \equiv \left\langle X\left(\{e\}, \{e^{\prime}\}\right) \right\rangle_{LHV} =$$ 
\begin{equation}
\label{xlhv}
\sum_{e_1, \ldots, e_n; e_1^{\prime}, \ldots, e_n^{\prime}} X\left(\{e\}, \{e^{\prime}\}\right)P\left(\{e\}, \{e^{\prime}\}\right),
\end{equation}
where
\begin{equation}
\label{xee'}
X\left(\{e\}, \{e^{\prime}\}\right) = \prod_{k = 1}^{n} \overline{e_k} - \prod_{k = 1}^{n} e_k^{\prime} + \sum_{i = 1}^{n} e_i\prod_{k = 1, k \ne i}^{n} e_k^{\prime}.
\end{equation} 
Note that $X\left(\{e\}, \{e^{\prime}\}\right)$ is non-negative if $\prod_{k = 1}^{n} e_k^{\prime} = 0$. On the other hand, if $\prod_{k = 1}^{n} e_k^{\prime} = 1$, we then have 
$$X\left(\{e\}, \{e^{\prime}\}\right) = \prod_{k = 1}^{n} \overline{e_k} - 1 + \sum_{i = 1}^{n} e_i,$$
which equals 0 if all $\overline{e_k}$'s are 1, and which is $\ge 0$ if at least one $\overline{e_k} = 0$. Thus $X\left(\{e\}, \{e^{\prime}\}\right)$ is $\ge 0$ for all values of the argument. Hence $X_{LHV} \ge 0$, \hfill{Q.E.D.}

{\noindent {\bf New results:}} We also obtain here a local reality upper bound on $X$ which extends the CH-Hardy inequality for $n = 2$ to general $n$. 

{\noindent {\underline{Upper bound on $X$:}}}
\begin{equation}
\label{upbonx}
X_{LHV} \le n - 1.
\end{equation}

{\noindent {\bf Proof:}} It suffices to show that $X\left(\{e\}, \{e^{\prime}\}\right) \le n - 1$ for all values of the arguments. If $\prod_{k = 1}^{n} e_k^{\prime} = 0$, at least one $e_k^{\prime}$, say $e_i^{\prime} = 0$. Then 
$$X\left(\{e\}, \{e^{\prime}\}\right) = \prod_{k = 1}^{n} \overline{e_k} + e_i\prod_{k = 1, k \ne i}^{n} e_k^{\prime} \le \prod_{k = 1}^{n} \overline{e_k} + e_i \le 1.$$
If $\prod_{k = 1}^{n} e_k^{\prime} = 1$, all $e_k^{\prime} = 1$, and hence
$$X\left(\{e\}, \{e^{\prime}\}\right) = \prod_{k = 1}^{n} \overline{e_k} - 1 + \sum_{i = 1}^{n} e_i.$$
If $\prod_{k = 1}^{n} \overline{e_k} = 1$, the right-hand side vanishes; if $\prod_{k = 1}^{n} \overline{e_k} = 0$, at least one $\overline{e_k}$, say $\overline{e_i} = 0$, {\it i.e.}, $e_i = 1$, and then
$$X\left(\{e\}, \{e^{\prime}\}\right) = \sum_{j = 1, j \ne i}^{n} e_j \le n - 1.$$
Thus $X\left(\{e\}, \{e^{\prime}\}\right) \le n - 1$ in all cases, with equality being reached when 
all $e_k^{\prime} = 1$, and all $e_i = 1$. \hfill{Q.E.D.}

\section*{\large{3. Chain of Hardy-type local reality inequalities}} The following identity provides a simple proof of 
the local reality inequality associated to Hardy's non-locality for general $n$, and suggests the existence of a chain
 of other local reality inequalities exhibiting Hardy-type non-locality. 

{\noindent {Master Identity:}} 
\begin{equation}
\label{mi1}
\prod_{k = 1}^{n} \overline{e_k} - \prod_{k = 1}^{n} e_k^{\prime} = \left(1 - \prod_{k = 1}^{n} e_k^{\prime}\right)\prod_{k = 1}^{n} \overline{e_k} - \left(1 - \prod_{k = 1}^{n} \overline{e_k}\right)\prod_{k = 1}^{n} e_k^{\prime},
\end{equation}
where
\begin{equation}
\label{mi2}
\left(1 - \prod_{k = 1}^{n} \overline{e_k}\right) = e_n + \sum_{i = 1}^{n - 1} e_i\prod_{j = i + 1}^{n} \overline{e_j}.
\end{equation}  
Equations (\ref{mi1}) and (\ref{mi2}) constitute the master identity. 

Our aim is to find bounds on the right-hand side of eq. (\ref{mi1}) in terms of $n$-fold products whose 
expectation values are experimentally measurable, i.e. they correspond to quantum expectation values of 
products of commuting observables. 
Using the fact that $\overline{e_j} \le 1$ on the right-hand side of equation (\ref{mi2}), and then using the fact that $e_k^{\prime} \le 1$, we get
\begin{equation}
\label{bonejek'}
 \left(1 - \prod_{k = 1}^{n} \overline{e_k}\right)\prod_{k = 1}^{n} e_k^{\prime} \le \sum_{i = 1}^{n} e_i\prod_{k = 1, k \ne i}^{n} e_k^{\prime}.
\end{equation} 
If instead we retain just one or two terms, $\overline{e_j}$ or $\overline{e_j}$  $\overline{e_k}$ in the coefficient 
of an $e_i$ in equation (\ref{mi2}), we get, successively
\begin{equation}
\label{ineq1}
\left(1 - \prod_{k = 1}^{n} \overline{e_k}\right)\prod_{k = 1}^{n} e_k^{\prime} \le e_i\overline{e_j}
\prod_{k = 1, k \ne i, j}^{n} e_k^{\prime} + \sum_{l = 1, l \ne i}^{n} e_l\prod_{k = 1, k \ne l}^{n} e_k^{\prime},
\end{equation}
for $i < j < n$, and 
$$\left(1 - \prod_{k = 1}^{n} \overline{e_k}\right)\prod_{k = 1}^{n} e_k^{\prime} \le e_i\overline{e_j}  
\overline{e_k}\prod_{l = 1, l \ne i, j, k}^{n} e_l^{\prime}$$ 
\begin{equation}
\label{ineq2}
+ \sum_{l = 1, l \ne i}^{n} e_l\prod_{k = 1, k \ne l}^{n} e_k^{\prime},
\end{equation}
for $i < j < k < n$, and a sequence of other inequalities by keeping more and more terms for the coefficient of 
an $e_i$ on the right-hand side of equation (\ref{mi2}). Of course one may also do that for the coefficients of 
more than one $e_i$ on the right-hand side of (\ref{mi2}), and obtain for example
$$\left(1 - \prod_{k = 1}^{n} \overline{e_k}\right)\prod_{k = 1}^{n} e_k^{\prime} \le e_i\overline{e_j}
\prod_{k = 1, k \ne i, j}^{n} e_k^{\prime} + e_k\overline{e_l}\prod_{m = 1, m \ne k, l}^{n} e_m^{\prime}$$
\begin{equation}
\label{ineq3}
+ \sum_{p = 1, p \ne i, k}^{n} e_p\prod_{q = 1, q \ne p}^{n} e_q^{\prime},
\end{equation}
for $i < j < k < l < n$, etc. Using the non-negativity of the first term on the right-hand side of equation (\ref{mi1}), we obtain the following chain of inequalities (by respectively using equations (\ref{bonejek'}), (\ref{ineq1}), (\ref{ineq2}), and (\ref{ineq3})), 
\begin{equation}
\label{chinq1}
X\left(\{e\}, \{e^{\prime}\}\right) = \prod_{k = 1}^{n} \overline{e_k} - \prod_{k = 1}^{n} e_k^{\prime} + \sum_{i = 1}^{n} e_i\prod_{k = 1, k \ne i}^{n} e_k^{\prime} \ge 0,
\end{equation}
$$X_{ij}\left(\{e\}, \{e^{\prime}\}\right) = \prod_{k = 1}^{n} \overline{e_k} - \prod_{k = 1}^{n} e_k^{\prime} + e_i\overline{e_j}\prod_{k = 1, k \ne i, j}^{n} e_k^{\prime}$$
\begin{equation}
\label{chinq2}
+ \sum_{l = 1, l \ne i}^{n} e_l\prod_{k = 1, k \ne l}^{n} e_k^{\prime} \ge 0~~ {\rm for}~ i < j < n, 
\end{equation}
$$X_{ijk}\left(\{e\}, \{e^{\prime}\}\right) = \prod_{l = 1}^{n} \overline{e_l} - \prod_{l = 1}^{n} e_l^{\prime} + e_i\overline{e_j} \overline{e_k}\prod_{l = 1, l \ne i, j, k}^{n} e_l^{\prime}$$
\begin{equation}
\label{chinq3}
+ \sum_{l = 1, l \ne i}^{n} e_l\prod_{m = 1, m \ne l}^{n} e_m^{\prime} \ge 0~~ {\rm for}~ i < j < k < n, 
\end{equation}
$$X_{ijkl}\left(\{e\}, \{e^{\prime}\}\right) = \prod_{m = 1}^{n} \overline{e_m} - \prod_{m = 1}^{n} e_m^{\prime} + e_i\overline{e_j}\prod_{m = 1, m \ne i, j}^{n} e_m^{\prime}$$
\begin{equation}
\label{chinq4}
+ e_k\overline{e_l}\prod_{m = 1, m \ne k, l}^{n} e_m^{\prime} + \sum_{p = 1, p \ne i, k}^{n} e_p\prod_{q = 1, q \ne p}^{n} e_q^{\prime} \ge 0 
\end{equation}
for $i < j < k < l < n$, and many others.

Multiplying successively equations (\ref{chinq1}) - (\ref{chinq4}) by $P(\{e\}, \{e^{\prime}\})$ and summing over all values of the $e_k$ and $e_k^{\prime}$, we obtain the inequality $X_{LHV} \ge 0$ stated before (which follows from equation (\ref{chinq1})) and the new inequalities
\begin{equation}
\label{newinq1}
\left(X_{ij}\right)_{LHV} \ge 0,~ i < j < n,~ n \ge 3,
\end{equation}
where
$$X_{ij} = P\left(e_k = 0~ \forall k\right) - P\left(e_k^{\prime} = 1~ \forall k\right) +$$
$$P\left(e_i = 1, e_j = 0, e_k^{\prime} = 1~ \forall k \ne i, j\right) +$$ 
\begin{equation}
\label{newinq1p}
\sum_{l = 1, l \ne i}^{n} P\left(e_l = 1, e_k^{\prime} = 1~ \forall k \ne l\right).
\end{equation}
Similarly,
\begin{equation}
\label{newinq2}
\left(X_{ijk}\right)_{LHV} \ge 0,~ i < j < k < n,~ n \ge 4,
\end{equation}
where
$$X_{ijk} = P\left(e_k = 0~ \forall k\right) - P\left(e_k^{\prime} = 1~ \forall k\right) +$$
$$P\left(e_i = 1, e_j = e_k = 0, e_l^{\prime} = 1~ \forall l \ne i, j, k\right) +$$ 
\begin{equation}
\label{newinq2p}
\sum_{l = 1, l \ne i}^{n} P\left(e_l = 1, e_k^{\prime} = 1~ \forall k \ne l\right),
\end{equation}
and
\begin{equation}
\label{newinq3}
\left(X_{ijkl}\right)_{LHV} \ge 0,~ i < j < k <  l < n,~ n \ge 5,
\end{equation}
where
$$X_{ijkl} = P\left(e_m = 0~ \forall m\right) - P\left(e_m^{\prime} = 1~ \forall m\right) +$$
$$P\left(e_i = 1, e_j = 0, e_m^{\prime} = 1~ \forall m \ne i, j\right) +$$
$$P\left(e_k = 1, e_l = 0, e_p^{\prime} = 1~ \forall p \ne k, l\right) +$$ 
\begin{equation}
\label{newinq3p}
\sum_{r = 1, r \ne i, k}^{n} P\left(e_r = 1, e_s^{\prime} = 1~ \forall s \ne r\right).
\end{equation}
Corresponding to the new lower bounds, given by equations (\ref{newinq1}), (\ref{newinq2}), and (\ref{newinq3}), 
we can also obtain the following upper bounds respectively. 

\vspace{0.2cm}
{\noindent {\underline{Upper bounds:}}}
\begin{equation}
\label{upbij}
\left(X_{ij}\right)_{LHV} \le n - 2,~~ {\rm for}~ n \ge 3; 
\end{equation}
\begin{equation}
\label{upbijk}
\left(X_{ijk}\right)_{LHV} \le n - 2,~~ {\rm for}~ n \ge 4; 
\end{equation}
\begin{equation}
\label{upbijkl}
\left(X_{ijkl}\right)_{LHV} \le n - 3,~~ {\rm for}~ n \ge 5. 
\end{equation}
For proofs of (\ref{upbij}) -- (\ref{upbijkl}), see the Appendix.

\section*{\large{4. New Hardy-type local reality constraints without inequalities}} Hardy's local reality constraints, 
corresponding to the condition $X_{LHV} \ge 0$, are already given by equations (\ref{ek'}) -- (\ref{ek}). In this 
section, we provide local reality constraints on joint probabilities corresponding to the conditions $(X_{ij})_{LHV} \ge 0$, 
$(X_{ijk})_{LHV} \ge 0$, and $(X_{ijkl})_{LHV} \ge 0$, without using these inequalities. 

\vspace{0.2cm}
{\noindent {(i)}} Suppose $P(e_k^{\prime} = 1~ \forall k) \ne 0$, and $P(e_i = 1, e_j = 0, 
e_k^{\prime} = 1~ \forall k \ne i, j) + \sum_{l = 1, l \ne i}^{n} P(e_l = 1, e_k^{\prime} =1 ~ \forall k \ne l) = 0$,
 $n \ge 3$. Then, $P(e_k = 0~ \forall k) \ge P(e_k^{\prime} = 1~ \forall k) \ne 0$.

\vspace{0.2cm}
{\noindent {\bf Proof:}} $P(e_k^{\prime} = 1~ \forall k \ne 0)$ implies that there exist events with $e_k^{\prime} = 1~ \forall k$; for these events, the vanishing of the probabilities under the summation means that $e_l = 0~ \forall l \ne i$, and in particular, $e_j = 0$; the vanishing of $P(e_i = 1, e_j = 0, e_k^{\prime} = 1~ \forall k \ne i, j)$ then means that $e_i = 0$, {\it i.e.}, all $e_k = 0$ for these events. Hence $P(e_k = 0~ \forall k) \ge P(e_k^{\prime} = 1~ \forall k) \ne 0$ which is a new Hardy-type local reality constraint that we proved without using the inequality $(X_{ij})_{LHV} \ge 0$. \hfill{Q.E.D.} 

\vspace{0.2cm}
Similarly we can prove the following new Hardy-type local reality constraints without using the associated Bell inequalities $(X_{ijk})_{LHV} \ge 0$, and $(X_{ijkl})_{LHV} \ge 0$.  

\vspace{0.2cm}
{\noindent {(ii)}} Suppose $P(e_k^{\prime} = 1~ \forall k) \ne 0$, and for $i < j < k < n$, 
$P(e_i = 1, e_j = e_k = 0, e_l^{\prime} = 1~ \forall l \ne i, j, k) + \sum_{l = 1, l \ne i}^{n} 
P(e_l = 1, e_k^{\prime} = 1 ~ \forall k \ne l) = 0$, $n \ge 4$. Then, $P(e_k = 0~ \forall k) \ge P(e_k^{\prime} 
= 1~ \forall k) \ne 0$.

\vspace{0.2cm}
{\noindent {\bf Proof:}} Since $P(e_k^{\prime} = 1~ \forall k) \ne 0$, there exist events with $e_k^{\prime} = 1$ for all $k$; the vanishing of the probabilities under the summation means that $e_l = 0$ for all $l \ne i$, and in particular $e_j = 0$ and $e_k = 0$ for these events; the vanishing of $P(e_i = 1, e_j = e_k = 0, e_l^{\prime} = 1~ \forall l \ne i, j, k)$ then implies $e_i = 0$ for these events. Hence 
$$P(e_k = 0~ \forall k) \ge P(e_k^{\prime} = 1~ \forall k) \ne 0,$$ 
a new Hardy-type local reality constraint. \hfill{Q.E.D.}

\vspace{0.2cm}
{\noindent {(iii)}} Suppose $P(e_k^{\prime} = 1~ \forall k) \ne 0$, and for $i < j < k < l < n$, $P(e_i = 1, e_j = 0, e_k^{\prime} = 1~ \forall k \ne i, j) + P(e_k = 1, e_l = 0, e_m^{\prime} = 1~ \forall m \ne k, l) + \sum_{p = 1, p \ne i, k}^{n} P(e_p = 1, e_q^{\prime} = 1~ \forall q \ne p) = 0$,~ $n \ge 5$. Then, proceeding as before, we can prove that 
$$P(e_k = 0~ \forall k) \ge P(e_k^{\prime} = 1~ \forall k) \ne 0,$$
another Hardy-type local reality constraint.

\vspace{0.2cm}
A chain of such constraints can be obtained by a straight forward extension of the simple argument (as mentioned above) 
without using the associated Bell inequalities.  

\section*{\large{5. Quantum violation of the new Hardy-type local reality constraints without inequalities}} 
In this section, we discuss quantum mechanical violations of the new Hardy-type local reality constraints 
without inequalities. Finding maximum possible violations of these constraints leads to optimization of the 
quantum states as well as of the choices of the observables $E_i$, $E_i^{\prime}$. The eigenstates $|e_i = 0\rangle$, $|e_i = 1\rangle$ of $E_i$ are related to the eigenstates $|e_i^{\prime} = 0\rangle$, $|e_i^{\prime} = 1\rangle$ of $E_i^{\prime}$ by a unitary transformation,
$$\left(\begin{array}{c}
|e_i^{\prime} = 0\rangle\\
|e_i^{\prime} = 1\rangle
\end{array}
\right) = \left(\begin{array}{cc}
- a_i & b_i\\
b_i^* & a_i^*
\end{array}
\right) \left(\begin{array}{c}
|e_i = 0\rangle\\
|e_i = 1\rangle
\end{array}
\right),$$
where $|a_i|^2 + |b_i|^2 = 1$. 

For $n = 3$, the maximum quantum mechanical violation of the Hardy-like constraints (\ref{ek'}) -- (\ref{ek}), 
corresponding to the CH-Hardy inequality (\ref{hardy-CH-Hardy}), is known to have the maximum possible value 
of the probability $P(e_k^{\prime} = 1~ \forall k)$ as $0.125$ \cite{ghosh}. We find here the maximum violation 
of the Hardy-like constraints in Sec. 4(i) for $n = 3$ explicitly, to illustrate the procedure. We seek a state         
$$|\psi\rangle = \sum_{e_1, e_2, e_3 \in \{0, 1\}} c_{e_1e_2e_3}|e_1e_2e_3\rangle$$
which maximizes $P(e_k^{\prime} = 1~ \forall k)$ by varying the coefficients $c_{e_1e_2e_3}$, subject to the 
constraints
$$P(e_k = 0~ \forall k) = 0,~ P(e_1 = 1, e_2 = 0, e_3^{\prime} = 1) = 0,$$
$$P(e_2 = 1, e_1^{\prime} = e_3^{\prime} = 1) = 0,~ P(e_3 = 1, e_1^{\prime} = e_2^{\prime} = 1) = 0,$$
and ${\langle}{\psi}|{\psi}{\rangle} = 1$. 

Since the constraints only involve probabilities in a single state, and not interference between different states, it turns out to be sufficent to choose phases of the states $|e_i\rangle$ such that the coefficients $c_{e_1e_2e_3}$ are real. Using the Lagrange method of undetermined multipliers we find that for $P(e_k^{\prime} = 1~ \forall k)$ to be stationary under variations of $c_{e_1e_2e_3}$, the ratios $b_j/a_j$ must be real, and we may choose the phases of the states $|e_j^{\prime} = 0\rangle$, $|e_j^{\prime} = 1\rangle$ such that the $a_j$ and $b_j$ are real. Stationarity is then achieved when the coefficients $c_{e_1e_2e_3}$ are chosen to be, $c_{000} = 0$ and 
$$\left\{c_{001}, c_{010}, c_{011}, c_{100}, c_{101}, c_{110}, c_{111}\right\} =$$ 
$${\pm}\frac{1}{\sqrt{(1 - b_1^2b_2^2)[b_3^2(1 - b_1^2b_2^2) + b_1^2b_2^2]}}\left\{b_3(1 - b_1^2b_2^2), a_2a_3b_1^2b_2,\right.$$ 
$$\left.- b_1^2a_2b_2b_3, a_1b_1b_2^2a_3, - a_1b_1b_2^2b_3, a_1a_2a_3b_1b_2, - a_1a_2b_1b_2b_3\right\}.$$ 

The stationary value of $P(e_k^{\prime} = 1~ \forall k)$ is then,
$$P(e_k^{\prime} = 1~ \forall k) = \frac{b_1^2b_2^2b_3^2(1 - b_3^2)(1 - b_1^2b_2^2)}{b_3^2(1 - b_1^2b_2^2) + b_1^2b_2^2},$$
which reaches a maximum value,
$$P(e_k^{\prime} = 1~ \forall k) = \frac{5\sqrt{5} - 11}{2} \approx 0.09017,$$
when
\begin{equation}
\label{b1b2b3}
b_1^2b_2^2 = b_3^2 = \frac{3 - \sqrt{5}}{2} = \left(\frac{\sqrt{5} - 1}{2}\right)^2.
\end{equation}
Since the solutions $b_1$, $b_2$, $b_3$ are not unique, the corresponding coefficients $c_{e_1e_2e_3}$, given above, are not unique, {\it i.e.}, the states $|\psi\rangle$ are not unique. A simple example is the choice $a_2 = 0$, $b_2 = 1$, $b_3 = b_1 = (\sqrt{5} - 1)/2$, $a_3 = a_1 = \sqrt{(\sqrt{5} - 1)/2} = \sqrt{b_1}$, which gives the state
$$|\psi\rangle = \sum_{e_1, e_2, e_3 \in \{0, 1\}} c_{e_1e_2e_3}|e_1e_2e_3\rangle,$$
with $c_{000} = c_{010} = c_{011} = c_{110} = c_{111} = 0$, and 
$$\left\{c_{001}, c_{100}, c_{101}\right\} = \frac{\sqrt{5} - 1}{2}\left\{1, 1, - \sqrt{\frac{\sqrt{5} - 1}{2}}\right\},$$

which achieves the maximum violation of the new Hardy-like constraint (described in sec. 4(i)) with 
$P(e_k^{\prime} = 1~ \forall k) = (5\sqrt{5} - 11)/2$. The state is not a generalized GHZ state. 
In fact, the state $|\psi\rangle$, given just above, is a tensor product of a single-qubit state and a 
two-qubit entangled state. Further, the continuum ambiguity in the solutions $b_1$, $b_2$, $b_3$, 
given by equation (\ref{b1b2b3}), translates to an one parameter family of state vectors $|\psi\rangle$ 
achieving the maximum violation $(5\sqrt{5} - 11)/2$.

\section*{\large{6. Cirel'son-type theorems on maximal violations of the 
corresponding CH-Hardy inequalities}} 
The usual Cirel'son theorem yields the maximum possible violations of the Bell-CHSH local reality inequality. Here we 
obtain analogous theorems on maximal quantum violations of the CH-Hardy inequalities for particular choices of 
the observables ${\bf X }$, ${\bf X_{ij}}$ whose expectation values are the corresponding measured quantities 
defined before. 

Let us first consider the quantum mechanical violation of the inequality (\ref{hardy-CH-Hardy}), where the 
quantity $X$ (given in equation (\ref{xee'})) is chosen to be the expectation value of the operator
\begin{equation}
\label{xop}
{\bf X }= \prod_{k = 1}^{n} \frac{I - {\sigma}_z^{(k)}}{2} - \prod_{k = 1}^{n} \frac{I + {\sigma}_x^{(k)}}{2} + 
\sum_{j = 1}^{n} H_j,
\end{equation}
with
\begin{equation}
\label{hj}
H_j = \frac{I + {\sigma}_z^{(j)}}{2}\prod_{k = 1, k \ne j}^{n} \frac{I + {\sigma}_x^{(k)}}{2},~ {\rm for}~ j = 1, 2, \ldots, n.
\end{equation}  
As $|\Phi\rangle \equiv \bigotimes_{k = 1}^{n} |z_-^{(k)}\rangle$, $|\chi\rangle \equiv \bigotimes_{k = 1}^{n}
 |x_+^{(k)}\rangle$, and $|{\Psi}_j\rangle \equiv |z_+^{(j)}\rangle \bigotimes_{k = 1, k \ne j}^{n} |x_+^{(k)}\rangle$ 
are eigenstates of the operators $\prod_{k = 1}^{n} (I - {\sigma}_z^{(k)})/2$, $\prod_{k = 1}^{n} (I + 
{\sigma}_x^{(k)})/2$, and $H_j$ respectively, we start with a `test' eigenstate of the operator ${\bf X}$ 
(in equation (\ref{xop})) as 
$$|\eta\rangle = {\alpha}|\Phi\rangle + {\beta}|\chi\rangle + {\gamma}|\Psi\rangle,$$
where the (unnormalized) state $|\Psi\rangle$ is given by $|\Psi\rangle = \sum_{j = 1}^{n} |{\Psi}_j\rangle$.   

It can be easily checked that 
$$ {\bf X}|\Phi\rangle = |\Phi\rangle - 2^{- n/2}|\chi\rangle,$$
$$ {\bf X} |\chi\rangle = 2^{- n/2}|\Phi\rangle - |\chi\rangle + 2^{- 1/2}|\Psi\rangle,$$
$$ {\bf X} |\Psi\rangle = - n \times 2^{- 1/2}|\chi\rangle + \left(\frac{n + 1}{2}\right)|\Psi\rangle.$$ 
Therefore, we have the follwing eigenvalue problem:
$$ {\bf X} |\eta\rangle = \left(\alpha + \frac{\beta}{(\sqrt{2})^n}\right)|\Phi\rangle - \left(\frac{\alpha}{(\sqrt{2})^n} + \beta + \frac{{\gamma}n}{\sqrt{2}}\right)|\chi\rangle$$
$$+ \left(\frac{\beta}{\sqrt{2}} + \frac{{\gamma}(n + 1)}{2}\right)|\Psi\rangle \equiv {\mu}|\eta\rangle.$$  
As the vectors $|\Phi\rangle$, $|\chi\rangle$ and $|\Psi\rangle$ are linearly independent, we then get from the above-mentioned eigenvalue equation:
\begin{equation}
\label{eigencoeff}
\begin{array}{lcl}
\alpha + 2^{- n/2}{\beta} &=& {\mu}{\alpha},\\
- 2^{- n/2}{\alpha} - {\beta} - \frac{{\gamma}n}{\sqrt{2}} &=& {\mu}{\beta},\\
\frac{\beta}{\sqrt{2}} + \left(\frac{{\gamma}(n + 1)}{2}\right) &=& {\mu}{\gamma}.
\end{array}
\end{equation} 
It follows from the first two conditions of equation (\ref{eigencoeff}) that 
\begin{equation}
\label{betagamma}
\begin{array}{lcl}
\beta &=& {\alpha} \times 2^{n/2}(\mu - 1),\\
\gamma &=& {\alpha} \times \frac{1}{n}\left[(1 - {\mu}^2)2^{(n + 1)/2} - 2^{(1 - n)/2}\right].
\end{array}
\end{equation}
Using these expressions into the last condition of equation (\ref{eigencoeff}), we get the follwoing cubic equation in the eigenvalue $\mu$, after cancelling out $\alpha$ on both sides of the condition (this is possible as $\alpha \ne 0$, otherwise, $\alpha = 0$ would give from equation (\ref{betagamma}) that $\alpha = \beta = \gamma = 0$, {\it i.e.}, the eigenstate $|\eta\rangle$ of $X$ is a null state):
\begin{equation}
\label{eigeneq}
2{\mu}^3 - (n + 1){\mu}^2 + {\mu}\left(2^{- n + 1} - 2 + n\right) - \left(n2^{- n} + 2^{- n} - 1\right) = 0.
\end{equation}

For a few small values of $n$, the (approximate) eigenvalues of ${\bf X}$ are given in Table I.  For $n = 2$, 
the eigenvalues are given by: $\mu = 1/2$, $(1 - \sqrt{2})/2 \approx - 0.20711$, $(1 + \sqrt{2})/2 \approx 1.20711$, 
while for $n = 3$, the approximate eigenvalues are given by: $\mu = - 0.22305$, $0.77294$, $1.4501$. 
Thus we see that the maximal violation of the CH-Hardy inequality (\ref{hardy-CH-Hardy}), for the choice 
of the observables $E_k = (I + {\sigma}_z^{(k)})/2$, $E_k^{\prime} = (I + {\sigma}_x^{(k)})/2$, is 
$- 0.20711$ (approx.) for $n= 2$, and $- 0.22303$ (approx.) for $n = 3$ . The eigenstates corresponding to 
these maximal violations are given by 
$${\alpha}\left[|\Phi\rangle + (\mu - 1)2^{n/2}|\chi\rangle + \frac{2^{(n + 1)/2}(\mu - 1)}{2\mu - n - 1}
|\Psi\rangle\right],$$
with ($n = 2, \mu \approx - 0.20711$), ($n = 3, \mu \approx - 0.22305$), ($n = 4, \mu \approx - 0.210496$), etc., 
where $|\Phi\rangle$, $|\chi\rangle$, and $|\Psi\rangle$ are defined above. Here, in the above-mentioned eigenstate, 
$\alpha$ is the normalization factor. Note that for $n = 2$, the eigenvalue $\mu =  (1 + \sqrt{2})/2 \approx 1.20711$ 
of the operator  ${\bf X}$  corresponds to a violation of the LHV upper bound $X \le n - 1$ by an amount $0.20711$ 
(approx.). But for $n = 3$, $n = 4$, etc., none of the three eigenvalues of $X$ gives rise to a violation of 
the LHV upper bound $X \le n - 1$. It should be mentioned here that we have listed the violations of the 
CH-Hardy inequality (\ref{hardy-CH-Hardy}) for a particular choice of the observables $E_k$, $E_k^{\prime}$. 
Other choices should be investigated separately.

\begin{table}
\caption{\label{tab:table1}Eigenvalues of ${\bf X}$  (given in equation (\ref{xop})) for different values of 
$n$ and the corresponding bounds from local realistic theory}

\vspace{0.2cm}
\begin{tabular}{|c|c|c|} 
\hline
{\it Values} & {\it Eigenvalues of} ${\bf X}$   & {\it LHV bounds} \\
{\it of} $n$ &                           &                   \\ \hline

2 & 1.20711,~ - 0.20711,~ 0.5 & 0 $\le X_{LHV} \le$ 1 \\ \hline

3 & 1.4501,~ - 0.223046,~ 0.77294 & 0 $\le X_{LHV} \le$ 2  \\ \hline

4 & 1.80652,~ - 0.210496,~ 0.903973 & 0 $\le X_{LHV} \le$ 3 \\ \hline

5 & 2.23266,~ - 0.190055,~ 0.957394 & 0 $\le X_{LHV} \le$ 4 \\ \hline

6 & 2.688752,~ - 0.1689639,~ 0.9802124 & 0 $\le X_{LHV} \le$ 5 \\ \hline

\end{tabular}
\end{table}

\vspace{0.3cm}
In the same fashion, let us now consider quantum mechanical violation of the LHV inequalities 
$0 \le X_{ij} \le n - 2$. Here we replace  $X_{ij}$ by the operator ${\bf X_{ij}}$, where
$$ {\bf X_{ij}} = \prod_{k = 1}^{n} \left(\frac{I - {\sigma}_z^{(k)}}{2}\right) - 
\prod_{k = 1}^{n} \left(\frac{I + {\sigma}_x^{(k)}}{2}\right) +$$ 

$$\left(\frac{I + {\sigma}_z^{(i)}}{2}\right)\left(\frac{I - {\sigma}_z^{(j)}}{2}\right)\prod_{k = 1, k \ne i, j}^{n} 
\left(\frac{I + {\sigma}_x^{(k)}}{2}\right)$$
\begin{equation}
\label{xijop}
+ \sum_{l = 1, l \ne i}^{n} \left(\frac{I + {\sigma}_z^{(l)}}{2}\right)\prod_{k = 1, k \ne l}^{n} \left(\frac{I + {\sigma}_x^{(k)}}{2}\right),~ i < j < n.
\end{equation} 

Following a similar technique as before,we seek eigenvectors $|\Theta\rangle$ and eigenvalues $\mu$ 
of ${\bf X_{ij}}$:  
\begin{equation}
{\bf X_{ij}} |\Theta \rangle = \mu |\Theta \rangle.
\end{equation}
We use an ansatz,
\begin{equation}
|\Theta\rangle = {\alpha}|\Phi\rangle + {\beta}|\chi\rangle + \gamma (|\Psi_{ij}\rangle + \sqrt{2} |\Psi_{j}\rangle) + \delta \sum_{l = 1, l \ne i,j }^{n} |\Psi_{l}\rangle,
\end{equation}
where $|\Phi\rangle $, $|\chi\rangle$ and $|\Psi_{j}\rangle$ have the same definitions as before, and 
\begin{equation}
|{\Psi}_{ij}\rangle \equiv |z_+^{(i)}\rangle |z_-^{(j)}\rangle \bigotimes_{k = 1, k \ne i,j}^{n} |x_+^{(k)}\rangle.
\end{equation}

We obtain, after long but straightforward calculations,
$${\bf X_{ij}} |\Theta \rangle = \left(\alpha + \frac{\beta}{(\sqrt{2})^n}\right) |\Phi\rangle$$
$$ - \left(\frac{\alpha}{(\sqrt{2})^n} + \beta + (3/2) \gamma +\frac{{\delta}(n-2)}{\sqrt{2}}\right)|\chi\rangle$$
$$+ \left(\frac{\beta}{\sqrt{2}} +\frac{3 \gamma}{2 \sqrt{2}}+ \frac{{\delta}(n - 1)}{2}\right) \sum_{l = 1, l \ne i,j }^{n} |\Psi_{l}\rangle$$
\begin{equation}
+ \left(\beta /2 + \gamma +\frac{(n-2) \delta}{2 \sqrt{2}} \right) (|\Psi_{ij}\rangle + \sqrt{2} |\Psi_{j}\rangle).
\end{equation}
This yields a quartic equation for the eigenvalues $\mu$ of $ {\bf X_{ij}} $,

$$\{4(\mu - 1)(2\mu - n + 1) - 3(n - 2)\}\{{\mu}(\mu - 1)$$ 
\begin{equation}
\label{xijeigen}
+ 2^{- n}\} + (4\mu - 1)(\mu - 1)(2\mu - 1) = 0. 
\end{equation}
For a few values of $n$, the (approximate) eigenvalues of $ {\bf X_{ij}} $ are given in Table II. For 
$n = 3$, the roots of equation ({\ref{xijeigen}) can be obtained exactly, and are given by: 
$\mu = (1 + \sqrt{1 + {\sqrt{3}}/2} )/2 \approx 1.183$, $(1 - \sqrt{1 + {\sqrt{3}}/2} )/2 \approx - 0.183$, 
$(1 + \sqrt{1 - {\sqrt{3}}/2} )/2 \approx 0.683$, $(1 - \sqrt{1 - {\sqrt{3}}/2} )/2 \approx 0.317$. 
Thus we see that the maximum eigenvalue $\mu = (1 + \sqrt{1 + {\sqrt{3}}/2} )/2$ corresponds to a 
violation of the upper bound $X_{ij} \le n - 2$ while the minimum eigenvalue $\mu = (1 - \sqrt{1 + {\sqrt{3}}/2} )/2$ 
corresponds to a violation of the lower bound $X_{ij} \ge 0$. For $n = 4, 5$, etc., none of the eigenvalues of 
$X_{ij}$ gives rise to a violation of the upper bound $X_{ij} \le n - 2$.

\begin{widetext}

\begin{table}
\caption{\label{tab:table2}Eigenvalues of $ {\bf X_{ij}}$ (given by equation (\ref{xijop})) for different values of 
$n$ and the corresponding bounds from local realistic theory}

\vspace{0.2cm}
\begin{tabular}{|c|c|c|} 
\hline
{\it Values of} $n$ & {\it Eigenvalues of} $ {\bf X_{ij}}$ & {\it LHV bounds} \\ \hline

3 & 1.183013,~ 0.6830127,~ - 0.1830127,~ 0.3169873 & 0 $\le (X_{ij})_{LHV} \le$ 1 \\ \hline

4 & 1.4667,~ 0.91912,~ - 0.19033,~ 0.30448 & 0 $\le (X_{ij})_{LHV} \le 2$ \\ \hline

5 & 1.911717,~ 0.9524242,~ - 0.1734191,~ 0.3092781 & 0 $\le (X_{ij})_{LHV} \le$ 3 \\ \hline

6 & 2.37600,~ 0.978917,~ - 0.156122,~ 0.301206 & 0 $\le (X_{ij})_{LHV} \le 4$ \\ \hline

7 & 2.8549439,~ 0.9847124,~ - 0.1447244,~ 0.3050681 & 0 $\le (X_{ij})_{LHV} \le$ 5 \\ \hline

\end{tabular}
\end{table}

\end{widetext}

\vspace{0.3cm}

\section*{\large{7. Conclusions}} The main result here is the discovery of a chain of local realistic inequalities 
involving joint probabilities for several two-level systems. Each of these inequalities may be considered 
as a multipartite generalization of the bipartite CH-Hardy inequality \cite{clauser, durt}. 
Corresponding to each of these inequalities, we have given new Hardy-type local realistic constraints 
without inequalities. These are generalizations of multipartite Hardy-type \cite{clifton} as well as 
Cabello-type \cite{cabello} constraints. An instance of quantum mechanical maximal violations of these 
constraints is given for a three-qubit system, and its value turned out to be $(5\sqrt{5} - 11)/2$ 
-- same as in the case of two qubits \cite{hardy92}. Interestingly, the three-qubit quantum state 
corresponding to the maximum violation is not unique -- it can have pure two-qubit entanglement or 
genuine three-qubit pure entanglement. This phenomenon is not seen in the case of maximum quantum 
mechanical violation of Hardy's non-locality constraints for three two-level systems, where this 
maximum violation is $1/8$ and the corresponding three-qubit states are locally unitarily connected 
to a three-qubit GHZ state \cite{ghosh}. A quantum violation of the CH-Hardy inequality (\ref{hardy-CH-Hardy}), 
corresponding to Hardy's non-locality constraint for three two-level systems (described by the conditions (\ref{ek'}) -- (\ref{ek})), is also given here. 

An interesting unresolved question is whether there are other interesting chains of CH-Hardy type inequalities. 
One may also think 
of extending our results to the case of multipartite multi-level systems. For a given $n$ ({\it i.e.}, 
for a given number of parties), the local realistic inequalities, described in this paper, are independent of 
each other as they were obtained by using independent conditions on the coefficients of 
$e_i$'s in equation (\ref{mi2}). One may try to compare these inequalities with the existing 
local-realistic inequalities involving joint probabilities \cite{chen} in the context of introduction of 
noise into the quantum states. On the other hand, towards obtaining more stringent CH-Hardy type inequalities for 
detection of multi-qubit entanglement, one may try to develop (instead of local reality), the separability bounds 
on the operators corresponding to $X$, $X_{ij}$, $X_{ijk}$, $X_{ijkl}$, etc. -- similar to what was done in 
ref. \cite{roy2005}.

\vspace{0.2cm}
\begin{center}
\large{\bf Acknowledgements}
\end{center}

\vspace{1cm}

S.M.R. is Raja Ramanna Fellow of the Department of Atomic Energy. He acknowledges discussions on quantum foundations  
with Monique Combescure and Jean-Marc Richard, collaborators in the project \# 3404 of the Indo-French Centre for 
promotion of advanced research (IFCPAR), and with Dipan K. Ghosh of IIT, Bombay.
S.M.R. and S.G. thankfully acknowledge the hospitality of the School of Physical Sciences, Jawaharlal Nehru University 
and that of the NIUS Camp (2010), held at the Homi Bhabha Centre for Science Education, where part this work was done.   

\section*{\large{Appendix}}

Here we provide the proofs of the three upper bounds, given by equations (\ref{upbij}) -- (\ref{upbijkl}).

Proof of $(X_{ij})_{LHV} \le n - 2$, for $n \ge 3$:-

{\noindent {\bf Proof:}} We start from 
\begin{equation}
\label{upbij1}
\left(X_{ij}\right)_{LHV} = \left\langle X_{ij}\left(\{e\}, \{e^{\prime}\}\right) \right\rangle_{LHV},
\end{equation} 
where $X_{ij}(\{e\}, \{e^{\prime}\})$ is defined by equation (\ref{chinq2}). We show that $X_{ij}(\{e\}, \{e^{\prime}\}) \in [0, n - 2]$ for all values of the $\{e\}$, $\{e^{\prime}\}$. If $\prod_{k = 1}^{n} \overline{e_k} = 1$, only the first two terms on the right-hand side of (\ref{chinq2}) can be non-zero, and then $X_{ij}(\{e\}, \{e^{\prime}\}) \in [0, 1]$. If $\prod_{k = 1}^{n} \overline{e_k} = 0$, at least one $\overline{e_k} = 0$, {\it i.e.}, either 

\vspace{0.2cm}
(i) $\overline{e_j} = 0$, or 

(ii) $\overline{e_j} = 1, \overline{e_i} = 0$, or 

(iii) $\overline{e_j} = 1, \overline{e_i} = 1$, and $\overline{e_m} = 0$ for some $m \ne i, j$. 
\vspace{0.2cm}

We evaluate $X_{ij}(\{e\}, \{e^{\prime}\})$ in these three cases separately.

\vspace{0.2cm}
{\noindent {{\underline{Case (i):}} For $\overline{e_j} = 0$, $X_{ij}(\{e\}, \{e^{\prime}\}) = \overline{e_j^{\prime}}\prod_{k = 1, k \ne j}^{n} e_k^{\prime} + e_j^{\prime}\sum_{l = 1, l \ne i, j}^{n} e_l\prod_{k = 1, k \ne l, j}^{n} e_k^{\prime}$ which lies in $[0, n - 2]$, since either $e_j^{\prime}$ or $\overline{e_j^{\prime}}$ must vanish, and there are then a maximum of $n - 2$ terms each belongs to $[0, 1]$.}} 

\vspace{0.2cm}
{\noindent {{\underline{Case (ii):}} For $\overline{e_j} = 1$, $\overline{e_i} = 0$, $X_{ij}(\{e\}, \{e^{\prime}\}) = (1 - e_i^{\prime}e_j^{\prime})\prod_{k = 1, k \ne i, j}^{n} e_k^{\prime} + e_i^{\prime}e_j^{\prime}\sum_{l = 1, l \ne i, j}^{n} e_l\prod_{k = 1, k \ne l, i, j}^{n} e_k^{\prime}$ which belongs to $[0, n - 2]$, since either $e_i^{\prime}e_j^{\prime}$ or $1 - e_i^{\prime}e_j^{\prime}$ must vanish.}}

\vspace{0.2cm}
{\noindent {{\underline{Case (iii):}} For $\overline{e_j} = 1$, $\overline{e_i} = 1$ and $\overline{e_m} = 0$ for some $m \ne i, j$, $X_{ij}(\{e\}, \{e^{\prime}\}) = \overline{e_m^{\prime}}\prod_{k = 1, k \ne m}^{n} e_k^{\prime} + e_m^{\prime}\sum_{l = 1, l \ne i, m}^{n} e_l\prod_{k = 1, k \ne l, m}^{n} e_k^{\prime}$ which belongs to $[0, n - 2]$, since either $e_m^{\prime}$ or $\overline{e_m^{\prime}}$ must vanish.}}

\vspace{0.2cm}
Finally $X_{ij}(\{e\}, \{e^{\prime}\}) \in [0, n - 2]$ for all values of the arguments with the upper bound being reached when all $e_k$ and all $e_k^{\prime}$ are equal to 1; hence $(X_{ij})_{LHV} \in [0, n - 2]$. \hfill{Q.E.D.} 

\vspace{0.2cm}
A similar calculation gives the upper bound
$$\left(X_{ijk}\right)_{LHV} \le n - 2,~ n \ge 4.$$
{\noindent {\bf Proof:}} If $\prod_{l = 1}^{n} \overline{e_l} = 1$, it then follows from (\ref{chinq3}) that $X_{ijk}(\{e\}, \{e^{\prime}\}) = 1 - \prod_{l = 1}^{n} e_l^{\prime}$, which lies in $[0, 1]$. If  $\prod_{l = 1}^{n} \overline{e_l} = 0$, we have either 

\vspace{0.2cm}
(i) $\overline{e_j} = 0$ or 

(ii) $\overline{e_k} = 0$ or 

(iii) $\overline{e_j} = \overline{e_k} = 1$, $\overline{e_i} = 0$ or 

(iv) $\overline{e_j} = \overline{e_k} = 1$, $\overline{e_l} = 0$ for some $l \ne i, j, k$. 

\vspace{0.2cm}
Let us now discuss these four cases separately.

\vspace{0.2cm}
{\noindent{\underline{Case (i):}} In this case, $X_{ijk}(\{e\}, \{e^{\prime}\}) = - \prod_{l = 1}^{n} e_l^{\prime} + \prod_{m = 1, m \ne j}^{n} e_m^{\prime} + \sum_{l = 1, l \ne i, j}^{n} e_l\prod_{m = 1, m \ne l}^{n} e_m^{\prime} = e_i^{\prime}\overline{e_j^{\prime}}\prod_{l = 1, l \ne i, j}^{n} e_l^{\prime} + e_i^{\prime}e_j^{\prime}\sum_{l = 1, l \ne i, j}^{n} e_l\prod_{m = 1, m \ne l, i, j}^{n} e_m^{\prime} \in [0, n - 2]$. 

\vspace{0.2cm}
{\noindent{\underline{Case (ii):}} Due to the symmetry of the two cases (i) and (ii) under $j \leftrightarrow k$, we again get, in this case, $X_{ijk}(\{e\}, \{e^{\prime}\}) \in [0, n - 2]$.

\vspace{0.2cm}
{\noindent{\underline{Case (iii):}} In this case, $X_{ijk}(\{e\}, \{e^{\prime}\}) = (1 - e_i^{\prime}e_j^{\prime}e_k^{\prime})\prod_{l = 1, l \ne i, j, k}^{n} e_l^{\prime} + e_i^{\prime}e_j^{\prime}e_k^{\prime}\sum_{l = 1, l \ne i, j, k}^{n} e_l\prod_{m = 1, m \ne l, i, j, k}^{n} e_m^{\prime} \in [0, n - 3]$.

\vspace{0.2cm}
{\noindent{\underline{Case (iv):}} $X_{ijk}(\{e\}, \{e^{\prime}\}) = (e_i^{\prime}e_j^{\prime}e_k^{\prime}\overline{e_l^{\prime}} + e_ie_l^{\prime})\prod_{q = 1, q \ne i, j, k, l}^{n} e_q^{\prime} + e_i^{\prime}e_j^{\prime}e_k^{\prime}e_l^{\prime}\sum_{m = 1, m \ne i, j, k, l}^{n} e_m\prod_{q = 1, q \ne i, j, k, l, m}^{n} e_q^{\prime} \in [0, n - 3]$.

\vspace{0.2cm}
Thus $X_{ijk}(\{e\}, \{e^{\prime}\}) \in [0, n - 2]$ in all cases with the upper bound being reached when all $e_k$ and all $e_k^{\prime} = 1$; hence $(X_{ijk})_{LHV} \le n - 2.$ \hfill{Q.E.D.}

\vspace{0.2cm}
A somewhat longer calculation yields the upper bound
$$\left(X_{ijkl}\right)_{LHV} \le n - 3,~ n \ge 5.$$
{\noindent{\bf Proof:}} If $\prod_{m = 1}^{n} \overline{e_m} = 1$, from equation (\ref{chinq4}), we then get that $X_{ijkl}(\{e\}, \{e^{\prime}\}) \in [0, 1]$. On the other hand, if $\prod_{m = 1}^{n} \overline{e_m} = 0$, either 

\vspace{0.2cm}
(i) $\overline{e_j} = 0$, or 

(ii) $\overline{e_l} = 0$, or 

(iii) $\overline{e_j} = \overline{e_l} = 1$, $\overline{e_i} = 0$, or 

(iv) $\overline{e_j} = \overline{e_l} = 1$, $\overline{e_k} = 0$, or (v) $\overline{e_j} = \overline{e_l} = \overline{e_i} = \overline{e_k} = 1$, and $\overline{e_p} = 0$ for some $p \ne i, j, l, k$. 

\vspace{0.2cm}
Let us now discuss these five cases separately.

\vspace{0.2cm}
{\noindent {\underline{Case (i):}}} In this case, $X_{ijkl}(\{e\}, \{e^{\prime}\}) = e_i^{\prime}e_k^{\prime}\overline{e_j^{\prime}}\prod_{q = 1, q \ne i, j, k}^{n} e_q^{\prime} + e_i^{\prime}e_k^{\prime}e_j^{\prime}[e_k\overline{e_l}\prod_{m = 1, m \ne k, i, j, l}^{n} e_m^{\prime} + e_l\prod_{q = 1, q \ne i, j, k, l}^{n} e_q^{\prime} + \sum_{p = 1, p \ne i, j, k, l}^{n} e_p\prod_{q = 1, q \ne p, i, j, k}^{n} e_q^{\prime}] \in [0, n - 3]$.

\vspace{0.2cm}
{\noindent {\underline{Case (ii):}}} This is similar to case (i) with the interchange $j \leftrightarrow l$, and hence $X_{ijkl}(\{e\}, \{e^{\prime}\}) \in [0, n - 3]$.

\vspace{0.2cm}
{\noindent {\underline{Case (iii):}}} In this case, $X_{ijkl}(\{e\}, \{e^{\prime}\}) = (e_i^{\prime}e_j^{\prime}e_k + (1 - e_i^{\prime}e_j^{\prime})e_k^{\prime}e_l^{\prime})\prod_{m = 1, m \ne i, j, k, l}^{n} e_m^{\prime} + \sum_{p = 1, p \ne i, j, k, l}^{n} e_p\prod_{q = 1, q \ne p}^{n} e_q^{\prime} \in [0, n - 3]$.

\vspace{0.2cm}
{\noindent {\underline{Case (iv):}}} This is similar to case (iii) with the interchange $i \leftrightarrow k$, and hence $X_{ijkl}(\{e\}, \{e^{\prime}\}) \in [0, n - 3]$.

\vspace{0.2cm}
{\noindent {\underline{Case (v):}}} In this case, $X_{ijkl}(\{e\}, \{e^{\prime}\}) = \overline{e_p^{\prime}}\prod_{q = 1, q \ne p}^{n} e_q^{\prime} + (\sum_{m = 1, m \ne i, j, k, l, p}^{n} e_m\prod_{q = 1, q \ne m, p}^{n} e_q^{\prime})e_p^{\prime} \le~ {\rm min} \{1, n - 5\}$.

\vspace{0.2cm}
Thus we see that in all these five cases, $X_{ijkl}(\{e\}, \{e^{\prime}\}) \in [0, n - 3]$, the maximum being reached when all $e_k = 1$, and all $e_k^{\prime} = 1$; hence $(X_{ijkl})_{LHV} \le n - 3$. \hfill{Q.E.D.}

\end{document}